\documentclass[linenumbers,preprint]{aastex631}

\usepackage{comment}
\usepackage{hyperref}
\usepackage{amsmath}
\usepackage{cleveref}
\usepackage{enumitem}
\usepackage{graphicx}
\usepackage{lmodern}

\newcommand{\Mnor}{\hbox{$\mathcal{M}^{\rm N}_\odot$}}

\newcommand{\Rnor}{\hbox{$\mathcal{R}^{\rm N}_\odot$}}

\received{November 13, 2022}
\revised{November 30, 2022}
\accepted{December 10, 2022}

\submitjournal{ApJL}

\shorttitle{Gamma Cas Stars}
\shortauthors{Gies et al.}
\graphicspath{{./}{figures/}}

\begin{document}
\nolinenumbers

\title{Gamma Cas Stars as Be + White Dwarf Binary Systems}

\correspondingauthor{Douglas R. Gies} \email{dgies@gsu.edu}

\author[0000-0001-8537-3583]{Douglas R. Gies}
\affiliation{Center for High Angular Resolution Astronomy and Department 
of Physics and Astronomy, Georgia State University, P.O. Box 5060, Atlanta,
GA 30302-5060, USA}

\author[0000-0003-4511-6800]{Luqian Wang}
\affiliation{Center for High Angular Resolution Astronomy and Department 
of Physics and Astronomy, Georgia State University, P.O. Box 5060, Atlanta,
GA 30302-5060, USA}
\affiliation{Yunnan Observatories, Chinese Academy of Sciences (CAS), Kunming 650216, Yunnan, China}

\author[0000-0002-4313-0169]{Robert Klement}
\affiliation{The CHARA Array of Georgia State University, Mount Wilson Observatory, Mount Wilson, CA 91023, USA}
\affiliation{ESO, Alonso de C\'{o}rdova 3107, Vitacura, Casilla 19001, Santiago de Chile, Chile}

\begin{abstract} 
\nolinenumbers

The origin of the bright and hard X-ray emission flux among the $\gamma$~Cas 
subgroup of B-emission line (Be) stars may be caused by gas accretion onto an 
orbiting white dwarf (WD) companion.  Such Be+WD binaries are the predicted 
outcome of a second stage of mass transfer from a helium star mass donor to a 
rapidly rotating mass gainer star.  The stripped donor stars become small and  
hot white dwarfs that are extremely faint compared to their Be star companions.  
Here we discuss model predictions about the physical and orbital properties of 
Be+WD binaries, and we show that current observational results on $\gamma$~Cas
systems are consistent with the expected large binary frequency, companion faintness 
and small mass, and relatively high mass range of the Be star hosts.  We determine 
that the companions are probably not stripped helium stars (hot subdwarf sdO stars), 
because these are bright enough to detect in ultraviolet spectroscopy, yet their 
spectroscopic signatures are not observed in studies of $\gamma$~Cas binaries.  
Interferometry of relatively nearby systems provides the means to detect very faint 
companions including hot subdwarf and cooler main sequence stars.  
Preliminary observations of five $\gamma$~Cas binaries with the CHARA Array 
interferometer show no evidence of the companion’s flux, leaving white dwarfs 
as the only viable candidates for the companions. 

\end{abstract}


\null
\vspace{1.0 cm}

\section{Introduction} \label{sec:intro}

B-emission line (Be) stars are rapidly rotating B-type stars that are losing 
gas and angular momentum into a circumstellar decretion disk \citep{Rivinius2013}. 
Some of these stars were spun-up through a prior stage of mass transfer in an 
interacting binary \citep{Pols1991,Hastings2021,Shao2021}, and the former donor star appears
as a stripped down remnant (He star, neutron star, or white dwarf).  Although 
faint, the remnants are detected as Be+sdO (hot subdwarf) systems through  
ultraviolet spectroscopy \citep{Wang2021} and as Be+NS systems as Be X-ray binaries
(BeXRBs; \citealt{Reig2011}). Most Be stars are modest X-ray sources with luminosities 
comparable to normal B-type stars \citep{Naze2022a}, but in addition to the BeXRBs, 
there is a subset of Be stars that have an unusually large and hard X-ray flux.  
This group is known as the $\gamma$~Cas stars or analogs, named after the bright prototype 
of the group \citep{Smith2016}.  There are several explanations for their 
bright X-ray flux.  The first envisions the creation of strong and localized 
magnetic fields near the star -- disk interface that can generate high 
temperature gas through flares \citep{Smith1998,Smith1999}.  The second idea
is that X-ray flux is generated by accretion onto a compact white dwarf (WD)
as occurs in cataclysmic variables \citep{Murakami1986,Hamaguchi2016}.  
Models of the X-ray spectrum for accretion onto a magnetic or non-magnetic WD
provide good fits of the X-ray spectra of $\gamma$~Cas and HD~110432 
\citep{Tsujimoto2018} and $\pi$~Aqr \citep{Tsujimoto2022}. 
\citet{Postnov2017} suggest a third scenario in which the companion is 
a rapidly spinning neutron star that avoids gas accretion through a   
``propeller'' mechanism, but \citet{Smith2017} argue that this idea fails to 
account for the number and observational properties of the $\gamma$~Cas group. 

The binary origin is partially motivated by the fact that several of the 
brightest $\gamma$~Cas stars are known binaries, including $\gamma$~Cas 
itself \citep{Nemravova2012} and $\pi$~Aqr \citep{Bjorkman2002}. 
Recently \citet{Naze2022b} presented the results of a radial velocity
survey of 16 $\gamma$~Cas stars that suggest that many are binaries. 
They found orbits for six previously unknown binaries and identified 
five other probable binaries that display significant velocity variability. 
These are all long period binaries with low mass companions whose flux 
was not detected.  These properties are expected for WD companions, but
\citet{Naze2022b} caution that these may also be systems with sdO companions
that generally are not X-ray bright \citep{Naze2022a}.  If so, then 
a binary companion alone is not a sufficient explanation for the X-ray 
properties of the $\gamma$~Cas stars, and the magnetic origin remains 
as an attractive explanation. 

Here we present several arguments that any sdO companions of the $\gamma$~Cas 
stars should be bright enough to detect through ultraviolet 
spectroscopy and near-IR interferometry. We suggest that the lack of any such 
detections to date indicates that the companions must be much fainter 
as expected for WD stars.  These considerations suggest that the 
$\gamma$~Cas stars probably have WD companions and that accretion onto 
the WD remains as a viable theory for their bright X-ray flux. 

\section{Formation Models} \label{sec:models}

The Be+sdO binaries were probably formed through Case~B mass transfer 
in which the donor filled its Roche surface during the expansion that 
occurred with the initiation of H-shell burning \citep{Pols1991,Shao2021}.
Most of the sdO stars represent the stripped-down remnants after 
completion of mass transfer, and they maintain their luminosity 
through He-core burning \citep{Gotberg2018}.  After a time comparable to
or less than the main sequence lifetime of the nearby Be star, the 
sdO stars will initiate He-shell burning and again grow in radius 
and luminosity \citep{Schootemeijer2018,Laplace2020}. 

This next stage of enlargement will lead to a second Roche-filling and 
mass transfer episode that is sometimes called Case~BB \citep{Delgado1981}. 
In general, if the remnant after this second mass transfer stage has 
a core mass greater than 1.4 \Mnor
\footnote{Nominal IAU solar units; \citet{Prsa2016}.}, 
then the star will quickly experience
advanced nuclear burning and explode as a H-poor supernova. In the lower 
mass case, nuclear burning will cease and the remnant will rapidly 
enter the WD cooling sequence as a CO WD (or ONe WD for larger masses;
\citealt{Dewi2002}).  

Detailed examples of this second mass transfer stage are given in 
studies by \citet{Habets1986} (see his Fig.\ 8) and by \citet{Willems2004} 
(see their Fig.\ 4 and Section 3.1.2).  Both of these models indicate
that the sdO star loses about $12\%$ of its mass through mass transfer 
to the Be star, and the post-mass transfer binary has a lower mass ratio
$M_2^\prime /M_1^\prime$ and a slightly longer orbital period.  \citet{Willems2004} 
present population statistics from their models of this evolutionary 
stage (that they call ``Channel 2'').  They find that the WD remnants 
have masses in the range 0.7 to 1.4 \Mnor ~and orbital periods in the 
range 40 to 1000 days (peaking at 200 days; see their Fig.\ 2, 
middle right panel).  The mass gainer (Be) stars end up with masses 
spanning the range of 7 to 17 \Mnor ~(see their Fig.\ 3, middle right panel).

The remnants with core mass below the Chandrasekhar limit will begin 
their lives as hot and small WDs shortly after the second mass transfer 
episode when nuclear burning ends.  \citet{Bedard2020} show examples of  
cooling curves for CO WDs for different ages and remaining H coverage 
(see their Fig.\ 6).  We list in Table~1 two examples of the probable  
stellar properties for remnant WD masses of 1.2 and 0.7 \Mnor .
The WD radii and temperatures are adopted from \citet{Bedard2020} for 
their H-thick envelope case and an age of 3.2 Myr (comparable to the 
ages of the more massive Be star companions).   We estimate the 
corresponding Be star masses by assuming conservative mass transfer
so that the post mass transfer mass ratio is 
$$ 
{{M_1^\prime} \over {M_2^\prime}} = 
{{M_1 + \triangle M_2} \over {M_2 - \triangle M_2}} =
{
{ {{M_1} \over {M_2}} + {{\triangle M_2} \over {M_2}} }  
\over 
{1- {{\triangle M_2} \over {M_2}} }
}.
$$
The mass ratio in the Be+sdO stage is approximately $M_1/M_2 = 8.0$
from  the recent compilation by \citet{Wang2023}, and we  assume a mass 
loss fraction in the second stage from the models cited above of 
${\triangle M_2} / M_2 = 12\% $.  The post mass transfer mass ratio is 
then approximately  $M_1^\prime / M_2^\prime= 9.2$, and we use 
this ratio to determine the mass of the Be star quoted in Table~1. 
We set the Be star temperature and radius from the main sequence (MS)  
mass calibration given by \citet{Pecaut2013}.  We also show in Table~1 
the parameters from \citet{Pecaut2013} for low mass MS stars
of the same mass as assumed for the WDs. 

\begin{deluxetable*}{lccccccc}[h]
\tablecaption{ Sample Be+WD System Parameters \label{tab1} }
\tablewidth{0pt}
\tablehead{
\colhead{}   & \multicolumn{3}{c}{High mass} && \multicolumn{3}{c}{Low mass} \\
                     \cline{2-4}  \cline{6-8} 
\colhead{Parameter} & \colhead{Be}  &   \colhead{WD}  &  \colhead{MS}  
                    && \colhead{Be}  & \colhead{WD}  &  \colhead{MS} 
}
\startdata
$M$ (\Mnor)                      &      11.1 &   1.2 &  1.2 &&    6.5 &   0.7 &   0.7  \\
$R$ (\Rnor)                      &       5.4 & 0.006 &  1.3 &&    3.9 & 0.012 &   0.7  \\
$T_{\rm eff}$ (kK)               &        25 &    57 &  6.2 &&     19 &    42 &   4.4  \\
$\triangle m$(1450 \AA ) (mag)   &         0 &  12.7 & 24.4 &&      0 &   9.8 &  36.6  \\
$\triangle m$(1.65 $\mu$m) (mag) &         0 &  13.7 &  4.8 &&      0 &  11.5 &   5.5  \\
\enddata
\end{deluxetable*}

In order to illustrate just how faint the WD companions are at the start 
of the WD cooling sequence, we created model spectral energy distributions 
for each case using model fluxes from 
TLUSTY for the WD (OSTAR2002; \citealt{Lanz2003})
and the Be star (BSTAR2006; \citealt{Lanz2007}) and 
from ATLAS9 \citep{Castelli2003} for the low mass main sequence case.
The stellar flux ratio $f_2/f_1$ was calculated as a function of wavelength 
by multiplying the model flux ratio by the square of the radius ratio.
Table~1 gives the resulting flux ratios expressed as a magnitude difference 
for the cases of the far ultraviolet spectrum (evaluated at 1450 \AA ) and 
of the near-infrared $H$-band (1.65 $\mu$m).  We find that a WD companion 
would be too faint to detect by current means in both bands ($\triangle m > 10$
mag), while a MS companion might be found in the near-infrared.  

We caution that the estimates in Table~1 do not account for the possible 
added flux from disks around both components.  Be star disks produce a 
continuum excess that grows with wavelength and may change the flux from the Be star 
and its decretion disk by as much as 0.3 mag in the $H$-band \citep{Touhami2010}.
This would increase the magnitude difference between the Be star and its companion 
by the same amount. 
On the other hand, flux from accreting gas surrounding the WD could increase 
the WD apparent brightness and cause a decrease in the magnitude difference.  
For example, there is a hot continuum flux component in the FUV spectra of some 
dwarf novae in quiescence that can make the system appear about 0.5 mag brighter 
than expected for the WD alone \citep{Long2005,Urban2006}.  However, adjustments of 
this size for the magnitude difference will not change the basic conclusion from 
Table~1 that the WDs are too faint to detect.

\section{Observational Tests} \label{sec:tests}

There are a number of tests and predictions that follow from the Be+WD scenario
for the $\gamma$~Cas stars that are particularly valuable given recent 
spectroscopic and interferometric investigations of Be stars in the group. 
Here we outline these tests and compare the expectations with 
observational results. The main predictions are given in italics. 

\vspace{1cm}

\subsection{Binary Frequency}
{\it All $\gamma$~Cas stars are long period binaries with low mass companions.} 
It is difficult to evaluate the binary status of Be stars in general 
and the $\gamma$~Cas stars in particular because the Doppler shifts 
of the Be stars are small compared to their rotationally-broadened 
line widths and because the orbital periods are long.  Thus, spectroscopic 
investigations are demanding and require high quality observations 
obtained over long time spans. In addition, it is important to account 
for selection effects introduced by random orbital inclination. 
Nevertheless, the seminal radial velocity study by \citet{Naze2022b} 
does indicate that a large fraction of the $\gamma$~Cas stars 
are binaries with low mass companions. Of the 26 $\gamma$~Cas stars 
known at this time \citep{Naze2020a,Naze2022a}, 10 are binaries 
with established orbits and 5 are candidate binaries \citep{Naze2022b}.  
The remaining stars generally have too few radial velocity measurements
to determine the binary status.  It is reasonable to conclude that 
most $\gamma$~Cas stars are binaries, but more work will be needed 
to determine if all members are binaries. 

\subsection{Magnitude Difference}
{\it The WD companions are very faint compared to the Be components.}
We showed sample cases of the expected magnitude differences in 
Table~1, and these show that the WD flux is probably not detectable 
by current observational methods.  On the other hand, a number 
of recent investigations demonstrate that the flux of  
brighter sdO companions can be directly observed across the spectrum,
so it should be possible to determine if the companions are helium stars
(sdO/sdB) or WDs. 
 
Recent studies by \citet{Wang2021,Wang2023} summarize the advances in 
far ultraviolet (FUV) spectroscopy that have led to the detection of the 
spectral signature of the hot sdO components in some 20 cases to date
with a magnitude difference around $\triangle m$(1450 \AA ) $\sim 3.5$ mag. 
Most of these sdO stars are in the relatively fainter stage of 
He-core burning (see \citealt{Wang2021}, Fig.\ 17), and this suggests 
that FUV spectroscopy should be able detect most of the sdO companions 
in Be+sdO binaries.  Figure~1 illustrates the expected magnitude difference
from the models of \citet{Gotberg2018} for the sdO star and from the 
associated main sequence star parameters \citep{Pecaut2013} for the 
Be star with masses estimated from $M_1/M_2 = 8.0$ \citep{Wang2023}.
The plotted magnitude difference as a function of sdO mass confirms that 
the predicted and observed sdO fluxes are above the detection limit for 
HST FUV spectroscopy, $\triangle m$(1450 \AA ) $< 4.4$ mag \citep{Wang2021}.

\begin{figure*}[h]
\gridline{\fig{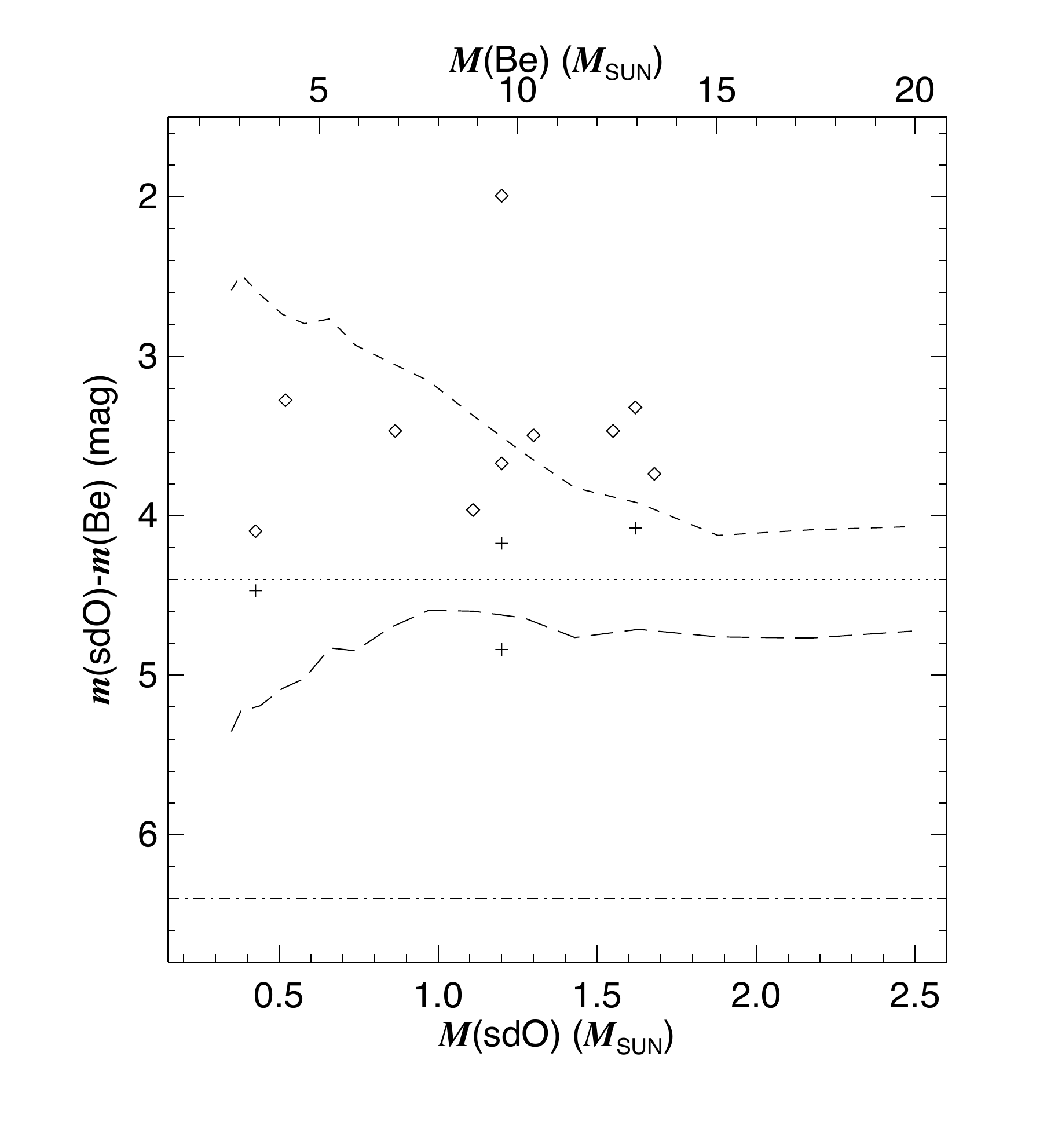}{0.75\textwidth}{} }
\caption{Predicted magnitude differences for Be+sdO binaries based upon the 
sdO flux models from \citet{Gotberg2018} for stars at the half-way point of He core 
burning.  The Be stars are assumed to have masses $8\times$ larger than the sdO stars 
and have temperatures and radii associated with main sequence values for the mass.
The upper, short dashed line shows the predicted magnitude difference at 
1450 \AA\ (FUV) while the lower, long dashed line is for 16500 \AA\ 
(near-infrared $H$-band).  Diamonds indicate observed values from FUV studies
and the crosses show those from CHARA Array $H$-band interferometry.
The FUV predictions and observations all are brighter than the faint limit 
of HST spectroscopy (middle dotted line), and the $H$-band models and 
observations are all much brighter than the detection limits of the CHARA Array
(lower dot-dashed line). These comparisons indicate that current observational 
methods are sufficient to detect the faint companion stars in most or all Be+sdO systems
(but not in the case of Be+WD binaries). }
\label{fig1}
\end{figure*}

\placefigure{fig1}

However, in contrast to the Be+sdO systems, all the FUV investigations to date of 
$\gamma$~Cas stars have led to null detections.  These include studies 
of IUE spectra of $\gamma$~Cas, $\pi$~Aqr, and $\zeta$~Tau \citep{Wang2017}, 
HD~45995 and HD~183362 \citep{Wang2018}, and of HST spectra of 
HD~157832 \citep{Wang2021}.  It is striking that no sdO component was 
found for these six binaries, but it is consistent with the idea that 
the companions of $\gamma$~Cas stars are faint WDs. Based upon their
expected faintness (Table~1), we predict that future HST FUV spectroscopy
will yield no detections of companions for the $\gamma$~Cas stars. 

The FUV results are confirmed in a recent visible-band spectroscopic investigation 
by \citet{Wang2023} who detected the presence of the \ion{He}{2} $\lambda 4686$
absorption line in the spectrum of the sdO star in four cases but found it was absent 
in the case of the $\gamma$~Cas star HD~157832. This again indicates that 
the companion star in HD~157832 is fainter than expected for an sdO star, 
and the lack of detection is consistent with a WD companion.  

Optical long baseline interferometry offers another means to investigate 
the companion flux of Be binaries that are close enough for their 
components to be angularly resolved.  
For example, sensitive closure phase measurements with the CHARA Array 
enable detection of companions with magnitude differences as large 
as $\triangle m$($1.65 \mu$m) = 6.4 \citep{Gallenne2015,Roettenbacher2015}.
CHARA Array observations have already successfully detected the companions 
and mapped the orbits of three Be+sdO and one Be+sdB binary systems  
\citep{Klement2022a,Klement2022b}, and a survey of other targets is 
underway.  If the companions in $\gamma$~Cas binaries are low mass MS stars, 
then these interferometric observations will detect them because they 
are relatively bright in the $H$-band (Table~1).  Furthermore, any 
sdO companions will also be bright enough for detection in the 
$H$-band (Fig.~1).  If, on the other hand, no companion flux is found, 
then the only remaining possibility is that the companions are faint WDs. 
Preliminary analysis of CHARA Array observations of five $\gamma$~Cas 
targets indicates no evidence of the flux of a companion (Klement et al., 
in preparation), which would rule out any sdO (Fig.~1) or MS (Table  1) 
companion. 

\subsection{Mass Ratio}
{\it The mass ratios of Be+WD binaries will be somewhat lower than 
those of Be+sdO systems.}
The second mass stripping episode will further decrease the remnant mass,
and models suggest that the proto-WD will lose about $12\%$ of its mass. 
It is difficult to determine the mass ratio in the absence of a 
double-lined spectroscopic orbital solution, but estimates are possible using the 
single-lined mass function plus estimates for the Be star mass and 
the orbital inclination. 
The estimated mass ratios for the well-studied $\gamma$~Cas stars are 
$q=M_2/M_1 = 0.075$ for $\gamma$~Cas \citep{Nemravova2012}, 0.086 for 
$\zeta$~Tau \citep{Ruzdjak2009}, 0.067 for HD~157832 \citep{Wang2023},
and 0.08 for $\pi$~Aqr \citep{Tsujimoto2022}.
These are all lower than the average value for Be+sdO systems of 
$q=0.125\pm 0.037$ \citep{Wang2023} as expected if the companions of the 
$\gamma$~Cas stars are lower mass WDs.  

Furthermore, the estimated companion masses are generally $<1\Mnor$, 
which is too low for a neutron star \citep{Fortin2016}. The one exception
is the binary $\pi$~Aqr, which \citet{Bjorkman2002} estimate has a 
companion of mass $2 \Mnor$. However, \citet{Naze2019} and 
\citet{Tsujimoto2022} have made new measurements of the Be star 
orbital velocity, and they both derive a semiamplitude that is $2\times$ 
smaller than that found by \citet{Bjorkman2002}.  This decrease leads 
to a smaller estimate of companion mass, $<1 \Mnor$, so the companion 
in the $\pi$~Aqr system does have a mass appropriate for a WD.  
\citet{Bjorkman2002} found a weak H$\alpha$ emission component
that displayed an antiphase radial velocity variation, and they 
suggested that the emission originates in the vicinity of the companion.  
This feature was conspicuous in spectra obtained between 1996 and 2000 when the 
H$\alpha$ disk emission was otherwise weak and confined to the extreme line wings, 
but subsequent spectra obtained when the disk was dense and large 
do not show this moving component \citep{Naze2019}.  The H$\alpha$ 
morphology described by \citet{Bjorkman2002} bears a strong resemblance
to that seen in the \ion{He}{1} $\lambda 6678$ line of the Be+sdO 
binary $\phi$~Per, and the antiphase motion in the latter case is 
explained as arising in a bright region of the Be star disk that 
faces and is illuminated by the hot companion \citep{Stefl2000,Hummel2001}.
\citet{Zharikov2013} find evidence of a similar bright region in 
the disk of $\pi$~Aqr that faces the companion. Thus, we suggest that
the emission component described by \citet{Bjorkman2002} probably
formed in the part of the Be disk directed towards the companion
and not in an accretion zone around the WD.  If so, then the mass 
ratio they derived from the emission velocity curve, $M_2/M_1=0.16$, 
is unreliable and is probably much larger than the actual value
($M_2/M_1=0.08 \pm 0.04$; \citealt{Tsujimoto2022}).

\subsection{Be Star Properties}
{\it The Be primaries in $\gamma$~Cas binaries will generally have high 
masses, very rapid rotation, and He-enriched atmospheres.} 
The second stage of mass transfer will probably add to the already 
large angular momentum of the mass gainer Be star, and the accreted 
gas may have a high proportion of nuclear processed He (and possibly C 
through the triple-$\alpha$ process).  The $\gamma$~Cas stars are 
generally found only among the more massive, early-type Be stars 
\citep{Smith2016,Naze2020a}, which is consistent with the mass range
of the gainer stars predicted by \citet{Willems2004} of 7 to 17 \Mnor . 
Furthermore, more massive Be stars will probably have more massive 
WD companions that have smaller radii, and outward leakage from the 
more massive Be star disks will yield a higher mass accretion rate by the WD. 
Both factors will result in a higher accretion luminosity from the WD. 

\pagebreak

\section{Conclusions} \label{sec:conclusions}

The strong and hard X-ray emission flux observed in the $\gamma$~Cas 
stars may result from gas accretion onto a small WD companion.  
Indeed, we argued here that the $\gamma$~Cas binaries represent the progeny
of the Be+sdO systems that are created after a second mass transfer stage 
that leads to the transformation of the helium star into a WD.  We discussed
the observational consequences of this Be+WD scenario for the $\gamma$~Cas 
stars, and we reviewed four broad tests of the validity of the model. 
The general observed properties are consistent with WD companions:
a large fraction of $\gamma$~Cas stars are known, long period binaries; 
the contrast in companion to Be star flux is too small to detect the 
companion's flux; the estimated mass ratios indicate companion masses 
below the Chandrasekhar limit; and the Be star hosts are very fast 
rotators at the high end of the Be star mass range. 
We demonstrated that the unseen companions in these systems are probably 
not subdwarf sdO stars, because the sdO stars are large and bright enough 
to detect their flux through analysis of ultraviolet spectroscopy,
yet none have been detected thus far among the $\gamma$~Cas stars. 
Near-infrared interferometric observations of targets that are near enough to resolve 
their orbits offer even more stringent limits on faint companions, and 
non-detection will rule out both helium star and main sequence star companions. 
Preliminary analysis of interferometry from the CHARA Array of five $\gamma$ Cas 
systems shows no evidence of the companion flux, and this leaves only WDs 
as viable companion candidates. 

The direct detection of such WD companions remains a daunting task 
because they are so much fainter than their Be star companions. 
It is possible that nulling interferometers \citep{Defrere2022} and 
coronagraphic imagers with extreme adaptive optics \citep{Davies2021} 
may be able to resolve and detect nearby Be+WD systems.  
Detection might also be possible through spectroscopic 
reconstruction methods based upon very high S/N and high resolving 
power observations that fully cover the orbital cycle.  For example, 
\citet{GiesRegulus2020} were able to detect the spectral signal from the WD 
companion of the B-star Regulus with a magnitude difference of $\triangle m$($V$) = 8.1 mag 
from a large set of high quality, visible wavelength spectra.  However, until 
such difficult observations are made, the best evidence for the WD companions 
of the $\gamma$~Cas stars is the consistent absence of their flux in the available 
observations because of their extreme faintness.

\vspace{0.5 cm}
This work is based upon observations obtained with the Georgia State University 
Center for High Angular Resolution Astronomy Array at Mount Wilson Observatory.  
The CHARA Array is supported by the National Science Foundation under Grant No. 
AST-1636624, AST-1908026, and AST-2034336.  Institutional support has been provided 
from the GSU College of Arts and Sciences and the GSU Office of the Vice President 
for Research and Economic Development. 
The work was also supported by the National Natural Science Foundation of China 
under programs Numbers 12103085, 12090040, and 12090043.
This research is also based on observations made with the NASA/ESA 
Hubble Space Telescope obtained from the Space Telescope Science Institute, 
which is operated by the Association of Universities for Research in Astronomy, Inc., 
under NASA contract NAS 5–26555. 
These observations are associated with program HST-GO-15659.
This work has made use of the SIMBAD database, operated at CDS, Strasbourg, France.

\vspace{5mm}
\software{TLUSTY \citep{Lanz2003,Lanz2007}, ATLAS9 \citep{Castelli2003} } 

\bibliography{ms.bib}{}
\bibliographystyle{aasjournal}

\end{document}